\documentclass[aps,prl,twocolumn,showpacs,superscriptaddress,amsmath,amssymb,citeautoscript]{revtex4}

\usepackage{graphicx}

\usepackage{color}

\usepackage{bm}% bold math

\begin{document}

\title{Topological order-by-disorder in orbitally degenerate dipolar bosons in zig-zag lattice}

\author{G. Sun}
\affiliation{Institut f\"ur Theoretische Physik, Leibniz Universit\"at Hannover, 30167~Hannover, Germany}
 \author {T. Vekua}
\affiliation{Institut f\"ur Theoretische Physik, Leibniz Universit\"at Hannover, 30167~Hannover, Germany}

\begin{abstract}
Spinor bosons offer conceptually simple picture of macroscopic quantum behavior of topological order-by-disorder:
Paramagnetic state of two-component dipolar bosons in orbitally degenerate zig-zag lattice is unstable against infinitezimal quantum fluctuations of orbitals towards developing non-local hidden order.
Adjacent to the topological state locally correlated exact ground state with spontaneously quadrupoled lattice constant is realized for the broad parameter regime.
\end{abstract}

\maketitle

\date{\today}

%\pacs{05.30.Fk, 03.75.Ss, 03.75.Mn, 71.10.Fd}

% Introduction

% REALIZATIONS AND MODEL

With the realization of the Mott insulator state of ultracold Bose gas loaded in optical lattice \cite{Greiner} a groundwork for experimental simulation of magnetism of many-body systems with bosons \cite{Lewenstein1,Giamarchi,Dalibard} was layed. Since then, with the help of shaking techniques, classical frustrated magnetism has been implemented successfully on triangular lattice \cite{Lewenstein}. 
Next target is to simulate quantum magnetism and in particular frustrated quantum spin systems to compensate for nonexistence of unbiased analytical or numerical methods and observe plaussible unconventional ground states a la spin liquids \cite{Sachdev, Balents}. Short-range quantum spin correlations for two-component alkali (contact interacting) Bose gases was exhibited in optical lattices \cite{Bloch}. Using bosonic dipolar atoms ($^{52}$ Cr with strong magnetic dipole moment) non-equilibrium quantum magnetism with long-range exchange physics has been reported in recent experiment \cite{Luis}. However, it needs a technological breakthrough in reducing temperatures below the spin coherence scales to simulate ground state equilibrium quantum magnetizm in experiments on ultracold gases \cite{deMarco}. 

Interestingly, lattice bosons can serve as well as an excellent {\it theoretical simulators} of a novel macroscopic quantum effect such as topological order-by-disorder that is possible to study by simple and at the same time solid analytical arguments.

To show this we study a system of 2-component dipolar bosons in orbitally degenerate zig-zag lattice depicted in Fig. 1. Due to an interplay between the geometric frustration caused by directional character and orbital degeneracy, and due to the bosonic nature arbitrary weak quantum fluctuations in orbitals select a tolopogical state from the manifold of extensively degenerate ground states. Adjacent to the topological state, for the broad regime of the system parameters, we also find an exact ground state of the product form with spontaneously broken translational symmetry having large unit cell made of 4 lattice sites.

\begin{figure}%[ht]
\begin{center}
\includegraphics[width=7.0cm]{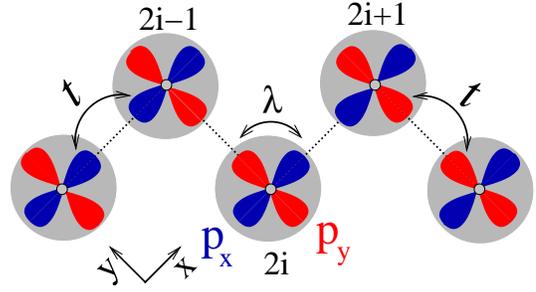}
\end{center}
\caption{Geometry of the orbitally degenerate zig-zag lattice with the nearest neighbour intersite hopping $t$ (between similar orbitals) and onsite hopping between the orthogonal orbitals $\lambda$.}
\label{fig:3}
\end{figure}

Ultracold bosons loaded in the degenerate $p$ bands of optical lattices have attracted a considerable theoretical and experimental interest \cite{Wu07,Wirth2011} due to possibility to observe chiral superfluid with emerging $p_x\pm ip_y$ order\cite{Olsh}.

Dipolar spinor bosons may be realized using diatomic polar molecules with electric dipole moment (e.g. potasium-rubidium $^{41}$K$^{81}$Rb\cite{Aikawa}) where spin-1/2 degrees of freedom may be encoded in two different total nuclear spin projection of molecules (similar to fermionic case $^{40}$K$^{81}$Rb \cite{Ospelkaus2010}), resulting in the fact that both long-range part as well as shot-range interactions will be largely spin independent.

First we derive the effective Hamiltonian describing the Mott insulator state of two-component dipolar bosons loaded in doubly degenerate $p$ bands of the zig-zag optical lattice where we retain two energetically degenerate orthogonal $p_{x}$ and $p_{y}$ orbitals per lattice site. The zig-zag lattice may be constructed by the incoherent superposition of a triangular lattice in $xy$ plane and an additional superlattice \cite{zigzag}. We assume that hopping between neighbouring sites is allowed only between the similar orbitals and amplitude of hopping we denote by $t$ as depicted in Fig. 1. 
Interactions between bosons occupying different sites (when orbitals are spatially separated) are to a good approximation orbital (and as already mentioned spin) independent, hence for deriving Hamiltonian describing the Mott phase corresponding to the average occupancy of one boson per lattice site it is sufficient to consider only interactions between bosons at the same site.

The interaction parameters for two bosons within
the same site are given by on-site repulsion within the same orbitals $U_{||}$ and between the orthogonal orbitals
$U_{\bot}$,
$
U_{||}(U_{\bot})=\int \mathrm{d} \mathbf {r_1} \mathrm{d} 
\mathbf {r_2}
  {p^2_{\alpha}}(\mathbf {r_1}   )   V(\mathbf {r_1}- \mathbf {r_2})
  {p^2_{\alpha(\beta\neq \alpha)}}( \mathbf {r_2} )
$.
Here $\alpha, \beta=x,y$, orbital wavefunctions $p_{x,y}(\mathbf {r})$ are assumed to be centered at the same site and $V(\mathbf {r_1}- \mathbf {r_2})$ is a total interparticle potential including both long-range and contact repulsive interactions.

Two bosons occupying the same orbital of one site may form an antisymmetric or a symmetric 
state with respect to the orbital index with corresponding energies 
$U_{||}\pm J_H$ which are split  by Hund's exchange $J_H=\int \mathrm{d} \mathbf {r_1} \mathrm{d} 
\mathbf {r_2}{p_x}(\mathbf{r_1})  {p_y}(\mathbf {r_1}) 
V(\mathbf {r_1}- \mathbf {r_2}){p_x}(\mathbf{r_2}) {p_y}(\mathbf {r_2}),
$ due to pair-hopping processes.
Two bosons occupying orthogonal orbitals of the same site may form singlet or triplet state in spin variables with corresponding energies $U_{\bot}\pm J_H$. In contrast to fermionic case, Hunds coupling minimizes total spin of bosons occupying orthogonal orbitals of the same site (even for contact repulsive interactions). This is due to minimization of repulsive interaction energy; By placing two bosons in antisymmetric $S^T=0$ spin singlet state bosonic nature demands the coordinate wavefunction to be antisymmetric as well, thus it has a node when distance between bosons vanishes and hence bosons avoid the region where repulsion would be the strongest.

In the strong coupling limit 
$U_{||}\pm J _H ,U_{\bot}\pm J_H \gg t$  and with one particle per site the system is in the Mott-insulator regime, and in second order perturbation theory in $t$ we arrive at the following spin-orbital model (SOM) Hamiltonian,
%%%%%%%%%%%%%
%\begin{eqnarray}
%\label{Themodel}
%H&=\!&-\sum_{i}^N[2 {\bf S}_i{\bf S}_{i+1}\!-\!\alpha+\!\frac{3}{2} ]
%\left[1\!+\!(-1)^i \sigma_i^z][1\!+\!(-1)^i \sigma_{i+1}^z\right]\nonumber\\
%&+&  \Delta \sum_{i}^N2{\bf S}_i{\bf S}_{i+1}
%\left[1- \sigma_i^z \sigma_{i+1}^z\right],
%\end{eqnarray}
%%%%%%%%%%%%%

%%%%%%%%%%%%%
\begin{eqnarray}
\label{Themodel}
H=&-&\sum_{i}(P_{i,i+1}+1- \alpha)
\left[1\!+\!(-1)^i \sigma_i^z][1\!+\!(-1)^i \sigma_{i+1}^z\right]\nonumber\\
&+&  \Delta \sum_{i}(P_{i,i+1}-1/2)
\left[1- \sigma_i^z \sigma_{i+1}^z\right],
\end{eqnarray}
%%%%%%%%%%%%%
where $P_{i,i+1}=2 {\bf S}_i{\bf S}_{i+1}+1/2$ is a permutation operator of spinor componenets expressed in terms of ${\bf S}_i$ spin-$\frac{1}{2}$ operators and $\sigma_i^{z}$ is a diagonal Pauli matrix describing the orbital
variables, with Eigenvalue $+1 (-1)$ corresponding to $p_x$ ($p_y$) orbital occupied on site $i$. We have fixed units of $t^2/2\tilde U=1$, with $\tilde U=(U_{||}^2-J_H^2)/U_{||}$, $\alpha=\tilde U (U_{\bot}-J_H/2)/(U_{\bot}^2-J_H^2)\sim U_{||}/U_{\bot}>0$ and $\Delta= J_H\tilde U/(U_{\bot}^2-J_H^2)>0$. Spin independence of interparticle interactions manifests in explicit SU(2) symmetry of the spin sector. %Note, despite the oscillating factor $(-1)^i$ in the first line of Eq. (\ref{Themodel}), the actual unit cell of the spin-orbital model is made of one site since renaming orbitals say at every odd site $p_x \leftrightarrow p_y$, $\sigma_i^z \to (-1)^i\sigma_i^z$ gives Hamiltonian (\ref{Themodel}) explicitely translationally invariant form.

We note here the crucial role of the long-range part of the interparticle interaction potential in deriving the SOM (\ref{Themodel}). For a purely contact interaction $V(\mathbf{r} )\sim \delta(\mathbf{r}) \to  U_{\bot}=J_H$, so that in the singlet spin channel 
two bosons located in the different orbitals of the same site do not experience any scattering.
Thus, when only $s$-wave contact scattering is present (typical case of alkali atoms), the Mott phase of one boson per lattice
well would be unstable due to orbital degeneracy. In the Mott-insulator regime, one can vary both $\alpha$ and $\Delta$ in a wide range by changing the lattice depth and a relative ratio of the strengths of the contact and dipolar interactions by modifying the dipole orientation by electric field or by tunning the contact interactions using Feshbach resonances.

\begin{figure}%[ht]
\begin{center}
\includegraphics[width=8.6cm]{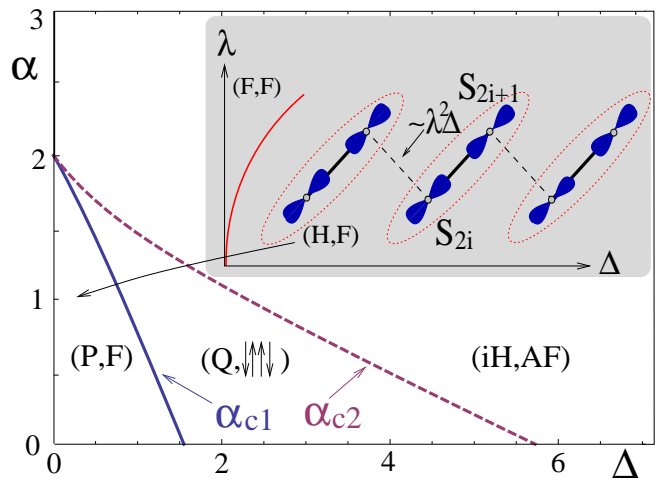}
\end{center}
\caption{Exact analytical ground state phase diagram of spin-orbital model Eq. (\ref{Themodel}) obtained in thermodynamic limit. We employ 
(spin,orbital) notation of different phases. For the ground state configurations of (Q,$\downarrow \uparrow \uparrow \downarrow$) and (iH,AF) phases see Fig. \ref{fig:Q} and for denotions of phases consult text.  Inset shows ground state orbital configuration of (P,F) phase and the effect in this phase of infinitezimal quantum fluctuations in orbitals $\lambda$. Dotted contours encircle 2 sites forming effective spins-$1$ ${\bf T}_i={\bf S}_{2i} +{\bf S}_{2i+1}$.}
\label{fig:GS}
\end{figure}

\begin{figure}%[ht]
\begin{center}
\includegraphics[width=7.5cm]{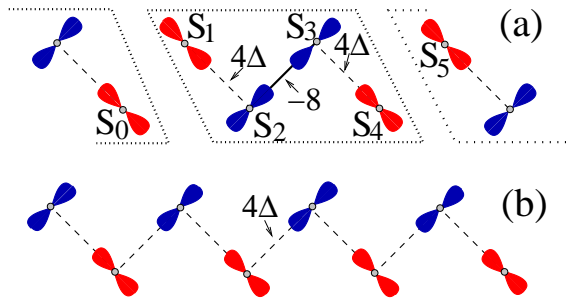}
\end{center}
\caption{Ground state orbital configurations in (a) (Q,$\downarrow \uparrow \uparrow \downarrow$) and (b) (iH,AF) phases. Only occupied orbitals are displayed per site. Dotted contour in (a) encircles cluster of 4 spins decoupled from the rest of the system. Continuous line indicates ferromagnetic Heisenberg exchange between spins at neighbouring sites  $-8 {\bf S}_i{\bf S}_{i+1}$ and dashed lines indicate AF exchange $4\Delta {\bf S}_i{\bf S}_{i+1}$.}
\label{fig:Q}
\end{figure}

{\it Ground-state phase diagram }--  Since orbital variables in Eq. (\ref{Themodel}) are classical it is easy to map out ground state phases in the product form of spin times orbital part. Depending on values of $\alpha$ and $\Delta$ only three different orbital configurations can be realized as ground states for Hamiltonian (\ref{Themodel}): a period of one ferromagnetic (F) as indicated in inset of Fig. \ref{fig:GS}, a period of two antiferromagnetic (AF) as depicted in Fig. \ref{fig:Q} (b), and a period of four configuration $\cdots p_yp_xp_xp_y \cdots$ ($\downarrow \uparrow \uparrow \downarrow $) presented in Fig. \ref{fig:Q} (a).

For $\alpha>2$ first line in Eq. (\ref{Themodel}) selects AF orbital configuration whereas the Hund coupling $\Delta$ induces AF exchange between the spins located on orthogonal orbitals of neighbouring sites. In spin sector one recovers isotropic Heisenberg antiferromagnet (iH) while in orbitals doubly degenerate AF configuration remains. Ground state energy per site in (iH,AF) state is independent of $\alpha$ and in the thermodynamic limit we can estimate it from an exact solution of spin-$\frac{1}{2}$ AF Heisenberg chain, $e^{iH}_0=\Delta(1-4\ln{2})$. There are no other phases for $\alpha>2$.

Phase diagram is much more interesting for $0<\alpha<2$ as presented in Fig. 2. There, besides (iH,AF) state we map out 2 additional ground states depending on $\Delta$ coupling.
For small values of $\Delta$ the ground state is two-fold degenerate and possesses F orbital order, 
$\langle \sigma_{i}^{z} \rangle=+1~(-1)$. Choosing $\langle \sigma_{i}^{z} \rangle=+1$ orbital configuration (this particular orbital order is selected by open boundaries if chain starts from even number site, see Fig. 1), two spins on the neighbouring sites combine to form an effective spin-1 $ {\bf T}_i={\bf S}_{2i}+{\bf S}_{2i+1} $ in the ground state, however ${\bf T}_i$ spins are completely decoupled from each other, thus resulting in extensively degenerate paramagnetic ground state (P) of spin-1 chain for the spin part of the wavefunction, with the total degeneracy of the ground state $2\times 3^{L/2}$ where $L$ is number of sites.

Energy per site for (P,F) configuration is $
e^P_0=2(\alpha-2)$ and is independent of $\Delta$.
Increasing $\Delta$ induces transition from (P,F) state into the ground state with $\downarrow \uparrow \uparrow \downarrow$ configuration of orbitals where bosons can hop only inside spontaneously selected 4 site clusters, as depicted in Fig. \ref{fig:Q}. For $\downarrow \uparrow \uparrow \downarrow$ configuration of orbitals coupling between the spins inside each decoupled cluster of 4 sites (see Fig. \ref{fig:Q}) is given by, $4\Delta{\bf S}_1 {\bf S}_2-8 {\bf S}_2 {\bf S}_3  +4\Delta  {\bf S}_3 {\bf S}_4+4\alpha-6$
and ground state energy per site is, $e^Q_0=\alpha-1-\frac{\Delta}{2}-\sqrt{1+\Delta+\Delta^2}.$ We denote this phase as (Q,$\downarrow \uparrow \uparrow \downarrow$) since spin exchanges have quadrumerized pattern.
Equating two energies $e^P_0=e^Q_0$
we obtain the phase transition line from (P,F) into (Q,$\downarrow \uparrow \uparrow \downarrow$) state, 
$\alpha=\alpha_{c1}= 3-\frac{\Delta}{2}-\sqrt{1+\Delta+\Delta^2}$ for any system size that is multiple of 4.

Further increasing $\Delta$ finally system minimizes its energy for (iH,AF) state since large $\Delta$, as already mentioned, induces antiferromagnetism for bosons. The phase transition line from (Q,$\downarrow \uparrow \uparrow \downarrow$) into (iH,AF) state is obtained by setting $e^Q_0= e^{iH}_0$ and is given in thermodynamic limit as,
$
\alpha=\alpha_{c2}=1+(3-4\ln4 )\Delta/2 +\sqrt{1+\Delta+\Delta^2}$. Different phases of bosons together with phase transition lines are presented in analytical phase diagram in Fig. \ref{fig:GS}.

In reality no optical lattice can be made ideally symmetric in $x-y$ plane, thus one can not neglect the nonzero probability of mixing of orbitals,
\begin{equation}
\label{QF}
H^{'}= - \sum_i  {\lambda}  \sigma_i^x,
\end{equation}
where $\lambda\ll t$ is infinitezimal perturbation. Alternatively, the $H^{'}$ perturbation can be controllably induced as an in-plane deformation of the 
lattice wells e.g. by an additional weak tilted lattice which leads to a mixing of the $p_{x,y}$ orbitals within the same well with an amplitude ${\lambda}$.

Recently, motivated by simulating transition-metal oxides 
with partially filled $d$-levels \cite{Konstantinovic04,Hikihara,Jackeli09}% such as pyroxene titanium oxides $\mathrm{ {\it A}TiSi_2O_6\, ({\it A}=Na,Li)}$~\cite{Konstantinovic04,Hikihara} that 
 containing zigzag chains of spin-$\frac{1}{2}$ ions, a similar to Eq. (1) SOM was introduced for fermions \cite{Sun+12} (with an essential difference of the overall sign in front of the Hamiltonian) and was shown that finite quantum fluctuations in orbitals can stabilize an exotic spin-orbital-liquid phase \cite{Sun+13}.

 The effect of arbitrary weak quantum fluctuation $\lambda$ on (P,F) state of bosons is remarkable: the perturbation $H^{'}$, acting as a transverse field in orbitals tries to quantum disorder orbital order in $\sigma^z$ variables that is otherwise perfect for $\lambda=0$, and at the same time, most importantly, it introduces exchange interactions between the decoupled neighbouring spins of (P,F) state,
\begin{equation}
\label{Hal}
H_S=-\sum_i 8 {\bf S}_{2i} {\bf S}_{2i+1}+\chi_{\lambda} \sum_i{\bf S}_{2i+1} {\bf S}_{2i+2},
\end{equation}
where $\chi_{\lambda}\simeq {\lambda^2 \Delta/2(1-\Delta)^2-O(\lambda^4)}$. In particular, in the limit $\lambda \to 0$ the two neighbouring spins ${\bf S}_{2i}$ and ${\bf S}_{2i+1}$ are coupled ferromagnetically with each other with the strength that is infinitely stronger than antiferromagnetic coupling  between ${\bf S}_{2i+1}$ and ${\bf S}_{2i+2}$. Hence for $\lambda \to 0$ ground state wavefunction of the spin part of Eq. (\ref{Hal}) coincides with the ground state of spin -1 chain \cite{Haldane} and a topological (H,F) state is established with non-local string order \cite{Nijs}. One can determine boundary of (H,F) state $\Delta_{F}\sim \lambda^2$ for $\Delta \to 0$. For $\Delta<\Delta_F$ fully polarized (F,F) state is selected for the ground state, and for $\Delta>\Delta_F$ (H,F) state is stabilized. Thus, for $\Delta>0$ infinitezimal quantum fluctuations $\lambda\to 0$ select from the extensively degenerate ground state manifold (P,F) a doubly degenerate state for periodic boundary conditions (degeneracy is due to orbital F order) and 4-fold degenerate state for open boundary conditions. As already mentioned open boundaries remove orbital degeneracy and hence the residual 4-fold degeneracy is purely due to the edge spins of the topological state. 

Extensive ground state degeneracy at classical level (similar to (P,F) phase for $\lambda=0$) is characteristic property of many frustrated spin systems \cite{Diep}. If degeneracy can be lifted either by thermal \cite{Villain, Chubukov} or by quantum fluctuations \cite{Rastelli, Henley} and as a result magnetic order developes such behavior is reffered as order-by-disorder. No unambiguous experimental confirmation of order-by-disorder has been reported in condensed matter magnetic systems, though there are suggestions to simulate it in experiments on ultracold spinor Bose gases \cite{Song, Turner}.
 Order by quantum disorder in orbitally frustrated electron system was predicted in two-dimensional square lattice \cite{Jackeli07}. Here we encounter with the emergence of topological order by quantum disorder in orbitally frustrated one-dimensional dipolar spinor bosons.

Other phases depicted in Fig. \ref{fig:GS} are stable with respect to infinitezimal quantum fluctuations in orbitals $\lambda$. In particular in (Q,$\downarrow \uparrow \uparrow \downarrow$) state the end spins of two adjacent decoupled (for $\lambda=0$) 4-spin clusters will get coupled due to $\lambda$ by AF exchange, i.e. the cluster on Fig. \ref{fig:Q} will be coupled to its neighbours by terms $\sim \Delta \lambda^2 ({\bf S}_{0}{\bf S}_{1}+  {\bf S}_{4}{\bf S}_{5})$.

\begin{figure}%[!H]
\begin{center}
\vspace{6pt}
\includegraphics[width=8.0cm]{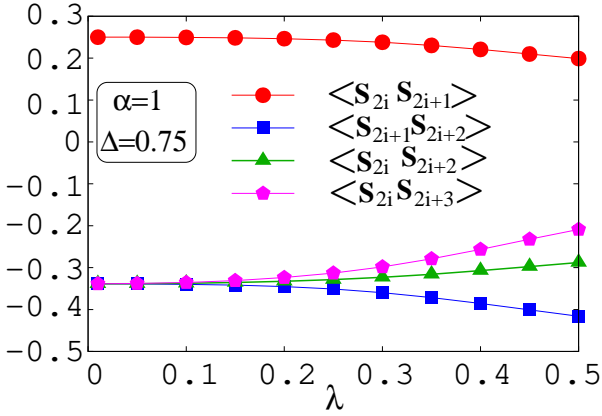}
\end{center}
\caption{ Bulk short-range spin correlation functions dependence on $\lambda$ in (H,F) phase. Due to extensive degeneracy of ground states in (P,F) phase, numerically we can not approach arbitrary close to $\lambda=0$, but the tendency is evident. Symmetry with respect to translations on 2 sites of (H,F) state imposes: $\langle {\bf S}_{2i}{\bf S}_{2i+1}\rangle = \langle {\bf S}_{2i+2}{\bf S}_{2i+3}\rangle$ and $\langle {\bf S}_{2i}{\bf S}_{2i+2}\rangle = \langle {\bf S}_{2i+1}{\bf S}_{2i+3}\rangle$ .}
\label{Cors}
\end{figure}

In the remaining we support our analytical findings numerically by simulating directly the full microsopic SOM including quantum fluctuations of orbitals, $H+H^{'}$. To address large systems we use density matrix renormalization group method~\cite{White} that is implemented best with open boundary conditions. Results of the numerical simulations of SOM presented below are for open system with $L=96$ sites and we compare them with the analogous results for the Haldane chain on $L=48$ sites to show that for $\lambda\to 0$ the ground state configuration of the spin part of the SOM reproduces identically topologically non-trivial ground state of the antiferromagnetic SU(2) symmetric spin-1 chain.

\begin{figure}%[ht]
\begin{center}
 \includegraphics[width=7.5cm]{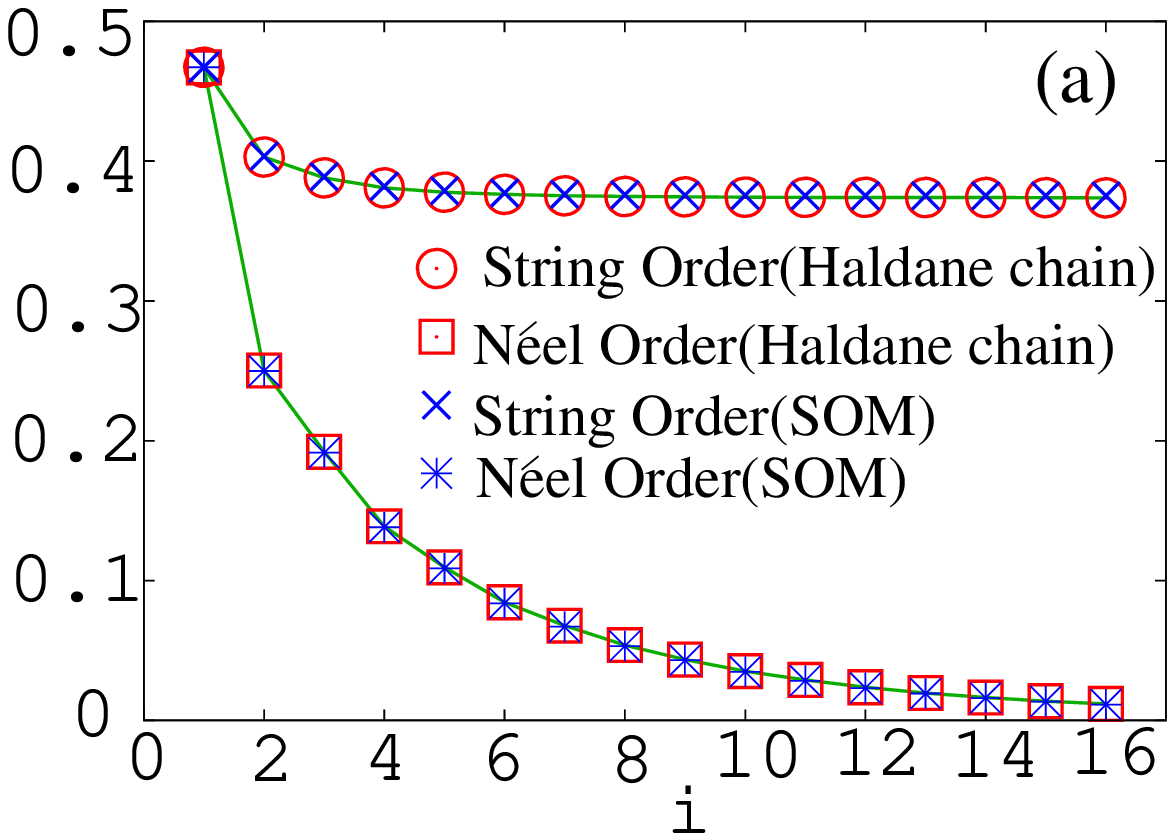} 
\includegraphics[width=7.9cm]{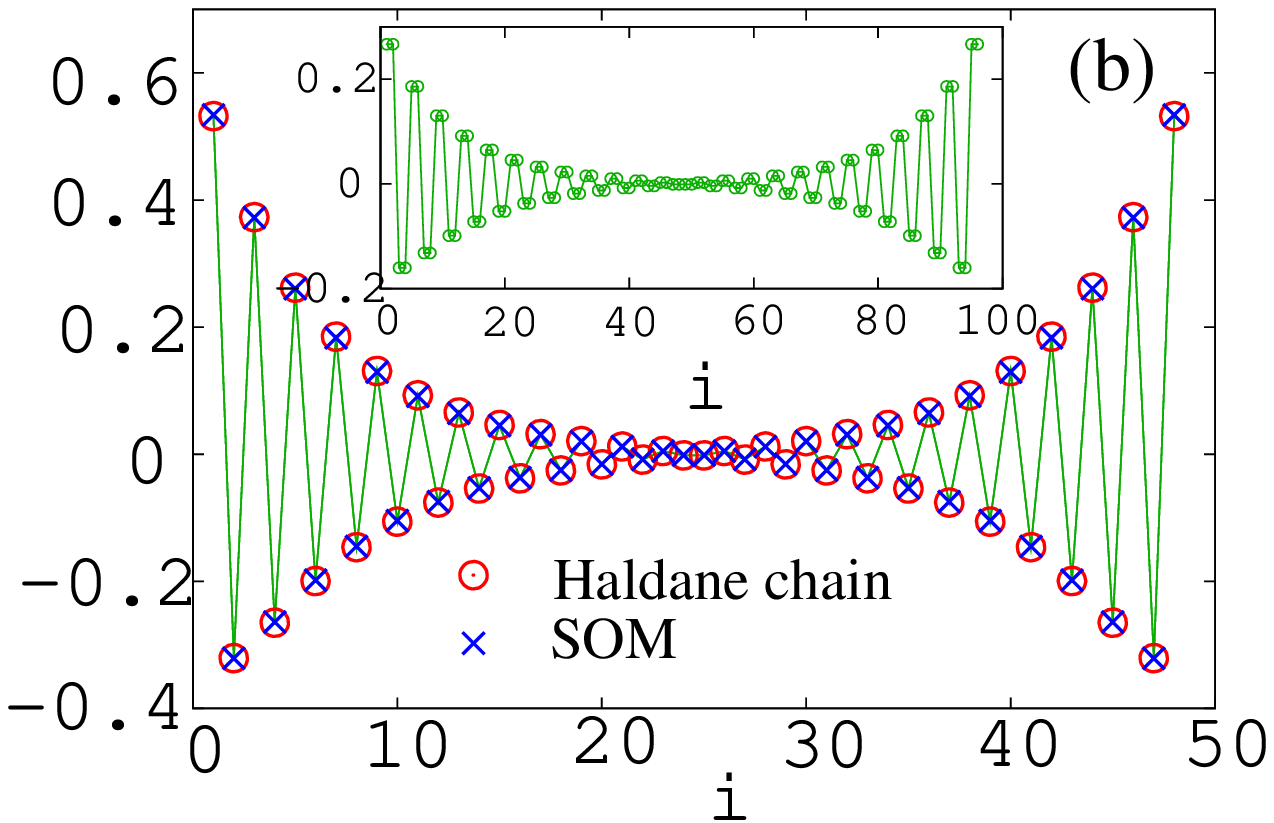} 
\end{center}
\caption{  (a)Blue symbols: bulk N\'eel order and string order of spin-orbital model (SOM) in (H,F) phase for $\lambda\to 0$ (here $\lambda=0.1$, $\alpha=1$ and $\Delta=0.1$ ). Red symbols: corresponding order parameters of spin-1 Haldane chain. (b) Magnetization profile in $S^z=1$ Kennedy-Tasaki ground state of spin-1 Haldane chain on $L=48$ sites (red symbols) is nearly identical to magnetization profile of $\langle T^z_i \rangle= \langle S^z_{2i}+S^z_{2i+1}\rangle  $ of SOM on $L=96$ sites (blue symbols) in (H,F) phase for $\lambda \to 0$ (here $\lambda=0.1$, $\alpha=1$ and $\Delta=0.1$). Inset (green circles) shows site resolved magnetization profile of SOM $\langle S^z_{2i}\rangle \simeq \langle S^z_{2i+1 }\rangle \simeq \langle T^z_{i}\rangle /2 $. }
\label{SN1}
\end{figure}

 First we present numerical results of short-range ground state correlation functions between the neighbouring spins as a function of $\lambda$ in (H,F) state in Fig. \ref{Cors}. As expected form analytical analyses one can observe in Fig. \ref{Cors} that in the limit $\lambda \to 0$: $\langle {\bf S}_{2i}{\bf S}_{2i+1}\rangle = \langle {\bf S}_{2i+2}{\bf S}_{2i+3}\rangle =1/4$ and $ \langle{\bf S}_{2i+1} {\bf S}_{2i+2}\rangle =\langle{\bf S}_{2i} {\bf S}_{2i+2}\rangle = \langle{\bf S}_{2i+1} {\bf S}_{2i+3}\rangle =\langle{\bf S}_{2i} {\bf S}_{2i+3}\rangle\simeq -0.35 $ so that $ \langle { \bf T}_i {\bf T}_{i+1}\rangle =\langle ({ \bf S}_{2i}+{ \bf S}_{2i+1}  ) ( {\bf S}_{2i+2} +{\bf S}_{2i+3}  )\rangle  \simeq -1.4 \simeq e_0(S=1)$ where $e_0(S=1)$ is the well known value of the ground state energy per site of the spin-1 chain~\cite{White} (in the units of exchange) that is equal to the ground state correlation function of two neighbouring spins of the Haldane chain.

N\'eel correlation function $(-1)^{j+i} \langle T^z_j T^z_{j+i} \rangle$ and string correlation function $- \langle T^z_j e^{i\pi \sum^{j+i-1}_{k=j+1} T^z_k}T^z_{j+i} \rangle$ are presented in Fig. \ref{SN1} (a) for both SOM on $L$ sites ($L=96$) and Haldane chain on $L/2$ sites. As one can see the coincidence between the results for the Haldane chain and SOM in $(H,F)$ state for small $\lambda$ is excellent.

Finally, magnetization profile of SOM $\langle {\bf S}_{2i}+{\bf S}_{2i+1}\rangle $ in one of the ground states of the Kennedy-Tasaki triplet\cite{KT} with total $S^z=1$ is presented in Fig \ref{SN1} (b). On the same plot we superimpose this profile with the corresponding profile of the Haldane chain \cite{White} and again observe the perfect matching between the two.

{\it In conclusion}, dipolar spinor bosons in orbitally degenerate zig-zag lattice develope topological order in extensively degenerate paramagnetic state due to arbitrary weak quantum fluctuations of orbitals . This is a direct consequence of the interplay between the orbital frustration and the bosonic nature. As a bonus adjacent to the topological state exact ground state is obtained with spontaneously quadrupoled unit cell for broad parameter regime of Hunds coupling and ratio between on-site and long range interactions.

We thank G. Jackeli for interesting us with spin-orbital models.
 This work has been supported by QUEST (Center for Quantum Engineering and 
Space-Time Research) and DFG Research Training Group
(Graduiertenkolleg) 1729.
%%%%%%%%%%%%%%
\\

\end{document}